\documentclass[aip,graphicx]{revtex4}

\pdfoutput=1

\usepackage{color}
\usepackage{graphics}
\usepackage[pdftex]{graphicx}
\usepackage{amsmath}
\usepackage{amssymb}
\usepackage{amsfonts}
\usepackage{float}
\usepackage{natbib}
\usepackage{epstopdf}
\usepackage{verbatim}
\usepackage{algorithm}
\usepackage{algorithmic}
\usepackage{lineno}
\usepackage{bbold}
\usepackage{array}
\usepackage{mathrsfs}
\usepackage{amsbsy}
\usepackage{mathtools}
\usepackage{chngcntr}
\usepackage{cleveref}

\crefname{equation}{}{}
\begin{document}

\title{Three-dimensional geometry and topology effects in viscous streaming}

\author{Fan Kiat Chan}
\affiliation{Mechanical Sciences and Engineering, University of Illinois at Urbana--Champaign, Urbana, IL 61801, USA}

\author{Yashraj Bhosale}
\affiliation{Mechanical Sciences and Engineering, University of Illinois at Urbana--Champaign, Urbana, IL 61801, USA}

\author{Tejaswin Parthasarathy}
\affiliation{Mechanical Sciences and Engineering, University of Illinois at Urbana--Champaign, Urbana, IL 61801, USA}

\author{Mattia Gazzola}
\email{mgazzola@illinois.edu}
\affiliation{Mechanical Sciences and Engineering, University of Illinois at Urbana--Champaign, Urbana, IL 61801, USA}
\affiliation{National Center for Supercomputing Applications, University of Illinois at Urbana--Champaign, Urbana, IL 61801, USA}
\affiliation{Carl R. Woese Institute for Genomic Biology, University of Illinois at Urbana--Champaign, Urbana, IL 61801, USA}

\begin{abstract}
Recent studies on viscous streaming flows in two dimensions have elucidated the impact of body curvature variations on resulting flow topology and dynamics, with opportunities for microfluidic applications.
Following that, we present here a three-dimensional characterization of streaming flows as function of changes in body geometry and topology, starting from the well-known case of a sphere to progressively arrive at toroidal shapes.
We leverage direct numerical simulations and dynamical systems theory to systematically analyze the reorganization of streaming flows into a dynamically rich set of regimes, the origins of which are explained using bifurcation theory.
\end{abstract}

\maketitle

\section{Introduction}
This paper investigates the role of body geometry and topology in three-dimensional viscous streaming settings.
Viscous streaming, a consequence of the non-linear nature of the Navier–Stokes equations, refers to the time-averaged steady flows that manifest when an immersed body of characteristic length $a$ is driven periodically with amplitude $A \ll a$ and frequency $\omega$ in a viscous fluid.
Streaming, which finds application in microfluidics for particle manipulation, trapping, sorting, assembly and passive swimming \citep{Liu:2002,Lutz:2003,Chung:2009,Tchieu:2010,Wang:2011,Chong:2013,Thameem:2016,Thameem:2017,Nair:2007,Klotsa:2015}, has been extensively studied and characterized theoretically, experimentally and numerically for constant curvature objects such as circular cylinders \citep{Holtsmark:1954,Riley:2001,Lutz:2005,Bhosale:2020}, infinite flat plates \citep{Glauert:1956,Yoshizawa:1974}, and spheres \citep{Lane:1955,Riley:1966,Kotas:2007}.
Beyond these uniform-curvature geometries, streaming flows involving objects of multiple curvatures received relatively little attention \citep{Badr:1994,Tatsuno:1975,Tatsuno:1974,Kotas:2007}, and studies have mostly focused on the observation and description of such flows, without establishing a mechanistic connection between shape geometry and flow reorganization.
In the pursuit of such explanation, recently, a systematic approach based on dynamical systems theory has been proposed in two-dimensional settings, revealing a rich set of novel flow topologies accessible via well-defined bifurcations, controlled through objects' local curvature and flow inertia \citep{Bhosale:2020}.
Expanded design space and rational design guidelines have then been elucidated to enhance existing applications or enable new ones, such as drug transport and delivery \citep{Parthasarathy:2019} by miniaturized swimming robots \citep{Park:2016,Aydin:2019,Ceylan:2017}.

In this work, we seek to extend this understanding to 3D settings.
We first consider simple axisymmetric flows, involving oscillating spheres and spheroids, to connect 2D insights to 3D observations.
We then depart from these simple cases and break flow axisymmetry by inverting spheroids' aspect ratios.
Since these configurations no longer have 2D analogues to guide our intuition, we analyze emerging flow topologies solely through a bifurcation theory perspective.
Finally, we explore the effect of body topology on streaming through the case of an oscillating torus and relate our observations to previously investigated spheroids of comparable length scales.

Overall, our study elucidates the mechanisms at play when three-dimensional streaming flow topology is manipulated through variations in objects' geometry, topology, and flow inertia,
thus providing physical intuition as well as design principles of potential use in microfluidics.

The work is organized as follows:
governing equations and numerical methods are summarized in $\S$\ref{sec:numerics};
streaming physics and flow topology classification are described in $\S$\ref{sec:background};
streaming flow characterization and transitions for axisymmetric flows and fully three-dimensional flows are investigated in $\S$\ref{sec:2d} and $\S$\ref{sec:3d}, respectively;
effects of body topology on viscous streaming flows are discussed in $\S$\ref{sec:topology};
finally, findings are summarized in $\S$\ref{sec:conclusion}.

\section{Governing equations and numerical method}\label{sec:numerics}
Here we briefly describe the governing equations and numerical techniques used in our simulations. We consider a solid body performing simple harmonic oscillations in an incompressible Newtonian fluid within an unbounded domain $\Sigma$. We denote the support and boundary of the density-matched solid with $\Omega$ and $\partial\Omega$, respectively. The three dimensional flow is then described by the incompressible Navier--Stokes equations
\begin{equation}
\boldsymbol{\nabla} \boldsymbol{\cdot} \boldsymbol{u} = 0, \qquad
\frac{\partial \boldsymbol{u}}{\partial t} + \left( \boldsymbol{u} \boldsymbol{\cdot} \boldsymbol{\nabla} \right)\boldsymbol{u} = -\frac{1}{\rho}\nabla p + \nu \boldsymbol{\nabla}^2 \boldsymbol{u}~~~~\boldsymbol{x}\in\Sigma\setminus\Omega
\label{eq:navier-stokes}
\end{equation}
where $\rho$, $p$, $\boldsymbol{u}$ and $\nu$ are the fluid density, pressure, velocity and kinematic viscosity, respectively.
Fluid--structure interaction is captured by solving equations \cref{eq:navier-stokes} in their velocity--vorticity form using remeshed vortex method, coupled with Brinkmann penalization to enforce the no-slip boundary condition $\boldsymbol{u} = \boldsymbol{u}_s$ at $\partial \Omega$, where $\boldsymbol{u}_s$ is the solid body velocity \citep{Gazzola:2011a}.
Our method has been validated across a range of fluid--structure interaction problems, from flow past bluff bodies to biological swimming and rectified flow phenomena \citep{Gazzola:2011a,Gazzola:2012a,Gazzola:2012,Gazzola:2014,Parthasarathy:2019,Bhosale:2020a,Bhosale:2020}.

\begin{figure}
\begin{center}
\includegraphics[width=\linewidth]{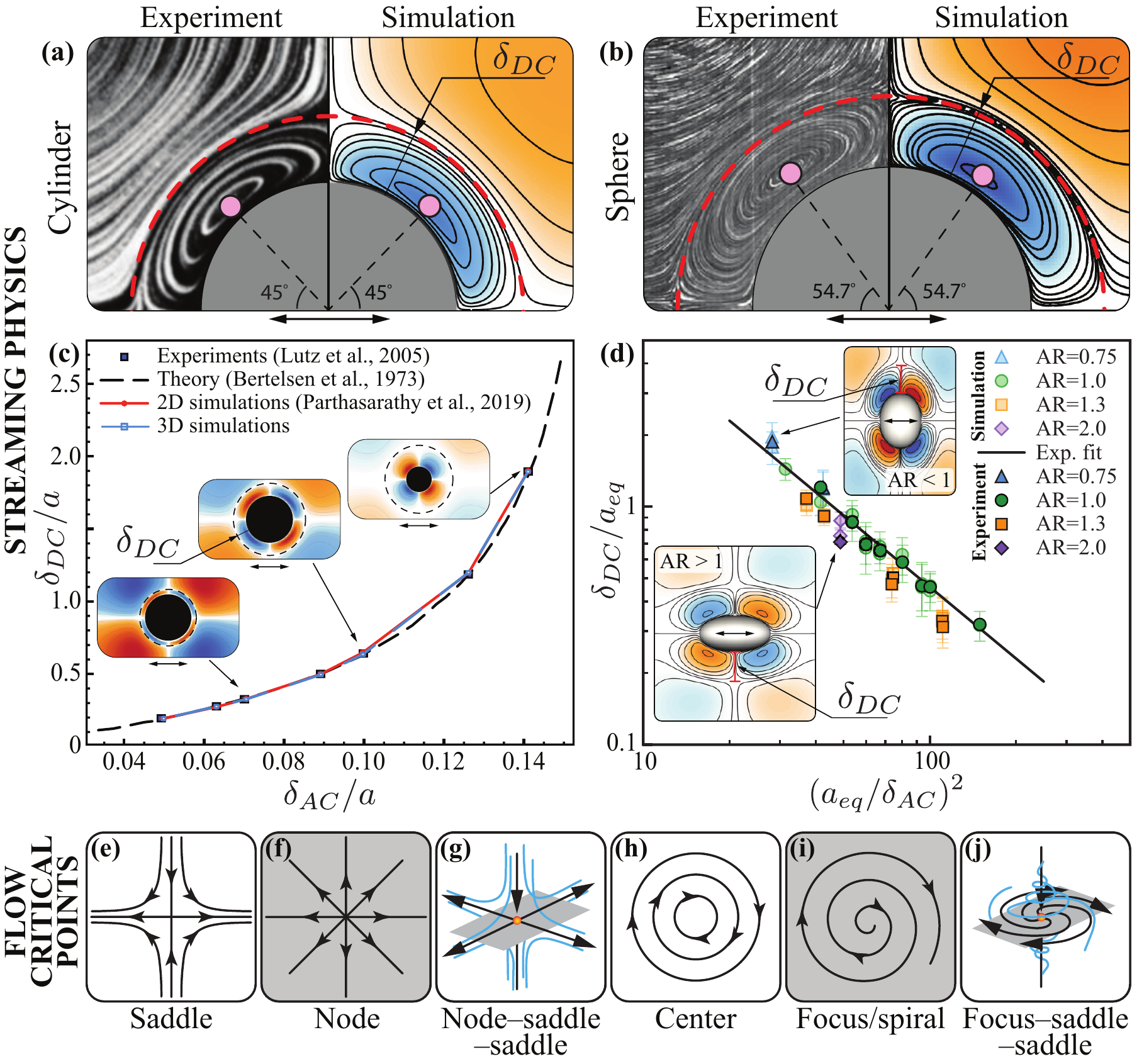}
\caption{Streaming physics: Comparison of time-averaged streamline patterns depicting regions of clockwise (blue) and counter-clockwise (orange) recirculating fluid for an oscillating \textbf{(a)} circular cylinder ($R_s = 0.628$, $\delta_{AC} / a = 0.126$) and \textbf{(b)} sphere ($R_s = 1.6$, $\delta_{AC} / a = 0.158$) in the finite thickness DC layer regime against experiments by \cite{vanDyke:1982} and \cite{Kotas:2007}, respectively.
The center of the inner vortex (pink marker) is observed at 45$^{\circ}$ and 54.7$^{\circ}$ from the axis of oscillation for the cylinder and sphere, respectively, consistent with the theory of \cite{Lane:1955}.
Quantitative comparison with experiments and theory in which we relate the normalized DC layer thickness $ \delta_{DC} / a$ and normalized AC boundary layer thickness $\delta_{AC} / a$ is illustrated in \textbf{(c)} for an oscillating cylinder \citep{Lutz:2005,Bertelsen:1973,Parthasarathy:2019} and \textbf{(d)} for oscillating spheroids of varying aspect ratios AR \citep{Kotas:2007}. For spheroids, the DC layer thickness (marked in red in inset) is defined as the average distance from the body surface to the stagnation points (saddles) perpendicular to the oscillation direction. The body length scale $a_{eq}$ for a spheroid is defined as the radius of an equivalent sphere of the same volume. The two-headed arrows in all subfigures indicate the direction of oscillation.
\textbf{(e--j)} Critical points and corresponding local flow patterns in three-dimensional incompressible flows.
Simulation details:
adaptive domain size with
uniform grid spacing $h = 1/2048$ (body length scale $a = 0.02$); penalization factor $\lambda = 10^4$; mollification length $\epsilon_{moll} = 2\sqrt{2}h$; Lagrangian Courant-Friedrichs-Lewy number $LCFL=0.01$; viscosity $\nu$ and oscillation frequency $\omega$ set according to prescribed $\delta_{AC} / a$. The above values are used throughout the text, unless stated otherwise. For more details on these parameters, we refer to \cite{Gazzola:2011a}.
}
\label{fig:validation}
\end{center}
\end{figure}

\section{Streaming physics and flow topology classification}\label{sec:background}
\subsection{Viscous streaming and numerical validation in two and three dimensions}\label{subsec:streaming}
We first introduce and characterize viscous streaming via the classical cases of circular cylinder and sphere of radii $a$. We consider a body immersed in a quiescent fluid of viscosity $\nu$ that performs low-amplitude harmonic oscillations defined by $x(t) = x(0) + A \sin(\omega t)$, where $A \ll a$ and $\omega$ are the dimensional amplitude and angular frequency, respectively.
The oscillatory motion then generates a Stokes layer of thickness $\delta_{AC} \sim \cal{O}(\sqrt{\nu/\omega})$ (commonly known as the AC boundary layer) around the solid body.
The velocity that persists throughout this layer drives a viscous streaming response in the surrounding fluid \citep{Batchelor:2000}.
Following \cite{Stuart:1966}, we characterize streaming response through the streaming Reynolds number $R_s = A^2 \omega / \nu$, based on the AC boundary layer thickness ($\delta_{AC} = A / \sqrt{R_s}$).
\Cref{fig:validation}(a,b) illustrates a comparison of time-averaged streamline patterns between our simulations and experiments \citep{vanDyke:1982,Kotas:2007} for circular cylinder and sphere at
$R_s = 0.628$ (or $\delta_{AC} / a = 0.126$) and $R_s = 1.6$ (or $\delta_{AC} / a = 0.158$), respectively.
The interplay between viscous and second-order inertial effects, for both circular cylinder and sphere, results in two classic flow topologies.
At high $R_s$ (or low $\delta_{AC} / a$), we encounter the double-layer regime characterized by a finite-thickness ($\delta_{DC}$) inner recirculating region (commonly known as DC layer) and an outer driven flow extending to infinity (\cref{fig:validation}(a,b)).
As $R_s$ decreases (or $\delta_{AC} / a$ increases), the inner region becomes thicker until it eventually diverges (\cref{fig:validation}(c)), extending to infinity and giving rise to the single-layer (or Stokes-like) regime.
For these simple shapes, there exist semi-analytical relations between the normalized AC layer ($\delta_{AC} / a$) and DC layer ($\delta_{DC} / a$) thicknesses \citep{Holtsmark:1954,Lane:1955,Bertelsen:1973}.
We then compare and validate our 3D simulations with theory, experiments and previous numerical investigations in \cref{fig:validation}(c).
As can be seen, both in this case, as well as against experiments involving oscillating spheroids (\cref{fig:validation}(d)), a quantitative match is obtained, thus verifying our 3D solver's accuracy.
We note that while in the simplest settings streaming dynamics is completely described by $\delta_{AC} / a$, this is not generally true when more complex shapes with multiple length scales are considered, and thus a more general approach becomes necessary.

\subsection{Flow topology characterization: dynamical systems theory}\label{subsec:flow-pattern}
Motivated by the need for a more generic approach to characterize streaming flows, we turn to dynamical systems theory, as previously proposed for 2D settings in \cite{Bhosale:2020}.
This approach offers a sparse yet complete representation of the underlying flow topology and its dynamics, and generalizes to 3D \citep{Chong:1990,Theisel:2003}.
We first identify the zero-velocity, critical points of the streaming field and classify these points based on their local flow properties, characterized through the eigenvalues/eigenvectors of the Jacobian $\mathbf{J}_{\boldsymbol{u}}$ associated with the velocity field \citep{Chong:1990}.
We recall that for a critical point, real components of the eigenvalues indicate local flow trajectories towards/away from (depending on the sign) the critical point, along the corresponding eigenvectors.
Imaginary components instead indicate rotational flows around the critical point, in the plane spanned by the corresponding eigenvectors.
For incompressible flows, where the trace of the Jacobian is always zero
(tr$(\mathbf{J}_{\boldsymbol{u}}) = \nabla \cdot \boldsymbol{u} = 0$),
only saddles (real eigenvalues of equal magnitude and opposite sign---\cref{fig:validation}(e)) and centers (imaginary eigenvalues of equal magnitude and opposite sign---\cref{fig:validation}(h)) exist in two-dimensional settings.
In three-dimensions, however, in-plane saddles and centers are accompanied by an out-of-plane component, which corresponds to the additional eigenvalue.
This allows for the existence of in-plane nodes (real eigenvalues of equal sign---\cref{fig:validation}(f)) and in-plane foci (complex-conjugate eigenvalues---\cref{fig:validation}(i)), both of which can be unstable/repelling or stable/attracting in nature, depending on the signs of the eigenvalues.
Under the incompressibility constraint, admissible combinations of local in-plane flows result in node--saddle--saddle (NSS, repelling example in \cref{fig:validation}(g)) and focus--saddle--saddle (FSS, repelling example in \cref{fig:validation}(j)) critical points \citep{Theisel:2003,Chong:1990}.

Following this characterization, we can then understand streaming flow reorganizations via bifurcation theory \citep{Strogatz:2018}, by analyzing the appearance and disappearance of critical points as shape features and flow inertia are modified. This allows us to systematically elucidate the mechanisms at play, to enable the rational manipulation of these systems.

\section{Axisymmetric flows}\label{sec:2d}
In order to understand the effects of geometry and topology variations on streaming dynamics in three dimensions, we consider first axisymmetric flows, which can be related back to more familiar two-dimensional settings \citep{Bhosale:2020}. This allows us to build intuition for interpreting more complex geometries and flows in later sections.
In axisymmetric cases, three-dimensional flow structures can be fully captured in a two-dimensional manner via the Stokes stream function \citep{Batchelor:2000}, and subsequently rendered in 3D using iso-surfaces for complete visual representation.
We note that while this approach provides natural intuition, it is only available under the condition of flow axisymmetry.
An alternative, compact and informative representation entails the extraction of critical points in combination with tracer particles.
These tracers can be seeded in the neighborhood of the critical points and then advected to reveal local flow features and connecting orbits, to further our physical intuition.
In the following, we analyze streaming flow structures through these two different perspectives, providing a comparison between a dense yet intuitive and a sparse yet complete flow representation.

\begin{figure}
    \begin{center}
        \includegraphics[width=\linewidth]{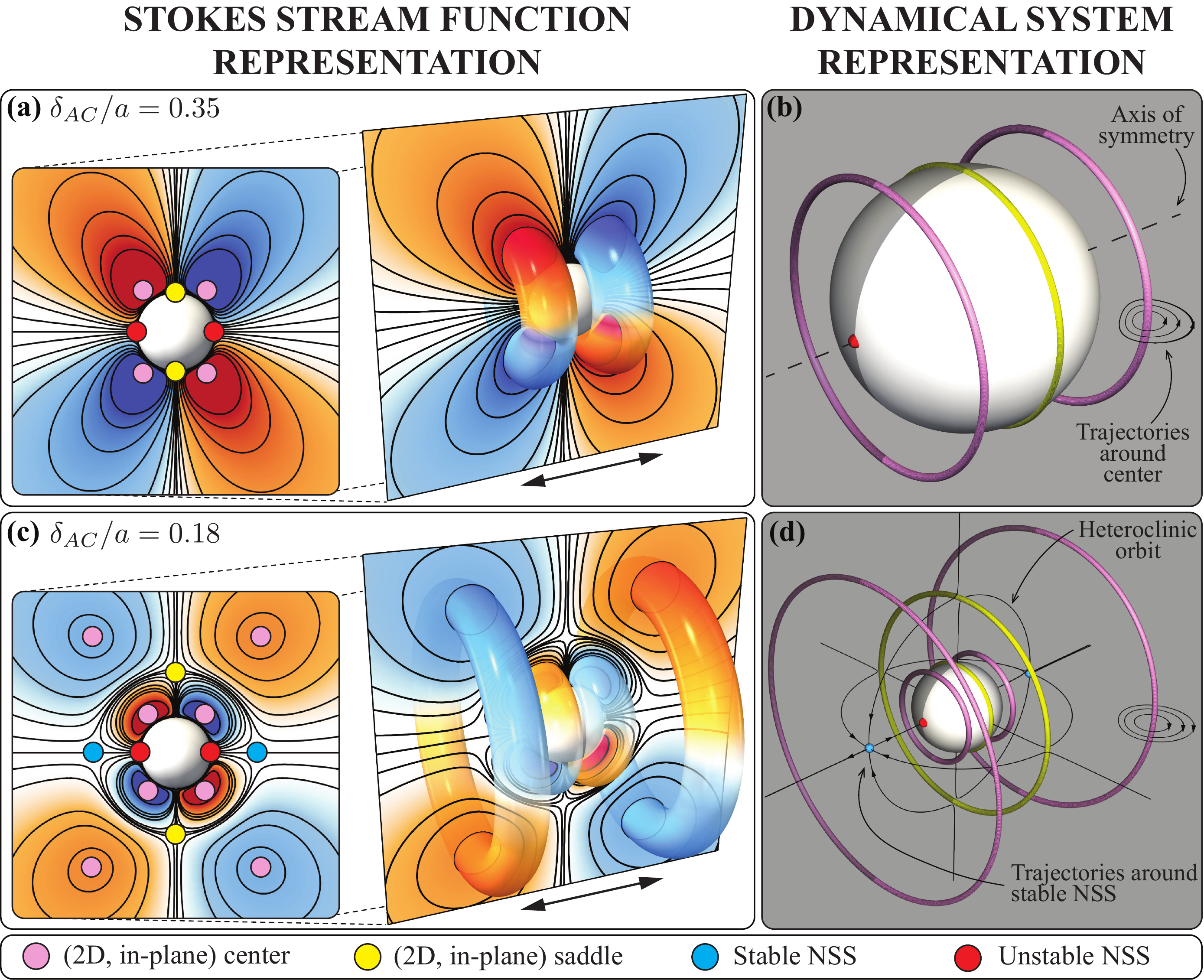}
        \caption{Streaming flow for a fully symmetric body (sphere of radius $a$) in the \textbf{(a,b)} Stokes-like ($\delta_{AC} / a = 0.35$) regime and \textbf{(c,d)} finite thickness DC layer ($\delta_{AC} / a = 0.18$) regime. The visualization of the flow field is presented using two different methods: Stokes stream function (left column) and dynamical system representation (right column). The 2D in-plane centers, 2D in-plane saddles, stable and unstable NSS (half-saddles/NSS on solid boundaries) are marked as pink, yellow, blue and red circles, respectively.
        Simulation details: normalized uniform grid spacing $h / a = 0.03$.
        }
        \label{fig:symmetric}
    \end{center}
\end{figure}

\subsection{Fully symmetric body: sphere}
We start by observing the streaming flows generated by an oscillating sphere (fully symmetric body) of radius $a$, initially in a quiescent fluid.

At high $\delta_{AC} / a$ (or low $R_s$), we encounter the Stokes-like regime. The corresponding Stokes stream function is illustrated in \cref{fig:symmetric}(a), where we observe the characteristic single-layer recirculating flow, highlighted via 3D iso-surfaces renderings.
The flow is organized around three types of critical points: centers (pink), saddles (yellow) and unstable NSS (red), marked here using circles.
We note that in this case both centers and saddles are 2D degenerate critical points \citep{Strogatz:2018}, since the surrounding local flow does not have any out-of-plane (i.e azimuthal) component due to axisymmetry.
These critical points, when mapped to three-dimensional space, form continuous rings as illustrated in \cref{fig:symmetric}(b), resulting in a flow skeleton that characterizes the system from a dynamical perspective.

As we decrease $\delta_{AC} / a$ (or increase $R_s$), we encounter the finite-thickness DC layer regime. In \cref{fig:symmetric}(c), we observe the characteristic double-layer recirculating flow, while in \cref{fig:symmetric}(d) we note the appearance of two additional outer center-rings (pink) as well as a saddle-ring (yellow), complemented by two new stable NSS (blue) that lie at a distance $\delta_{DC}$ away from the sphere surface.
Tracer trajectories further highlight the existence of heteroclinic orbits (trajectories connecting two different critical points) between the stable NSS and the degenerate saddle-ring, collectively forming a continuous spherical surface that cleanly separates the DC layer from the external driven fluid.

We note that due to axisymmetry, the transition between the single- and double-layer regime observed here in 3D for a sphere relies on mechanisms similar to the 2D circular cylinder \citep{Bhosale:2020}.
In the latter, the transition is mediated by higher-order reflecting umbilic bifurcations \citep{Bhosale:2020} for which, at a critical $\delta_{AC} / a$, saddles that are located at infinity split apart, eventually forming the outer centers as well as the saddles that delineate the DC layer (such process can be visualized in periodic domains, as demonstrated in \cite{Bhosale:2020}).
The details of the mechanisms through which critical points of various nature emerge from infinity are rather involved, and not particularly relevant to the remainder of our analysis.
Hence, for brevity, throughout the rest of the paper we refer this discussion to the Supplementary Information, while we focus instead on novel flow reorganizations observed in the proximity of the streaming body.

After briefly introducing our analysis procedure for the well-known case of the sphere, we proceed by progressively breaking symmetry.

\begin{figure}
    \begin{center}
        \includegraphics[width=\linewidth]{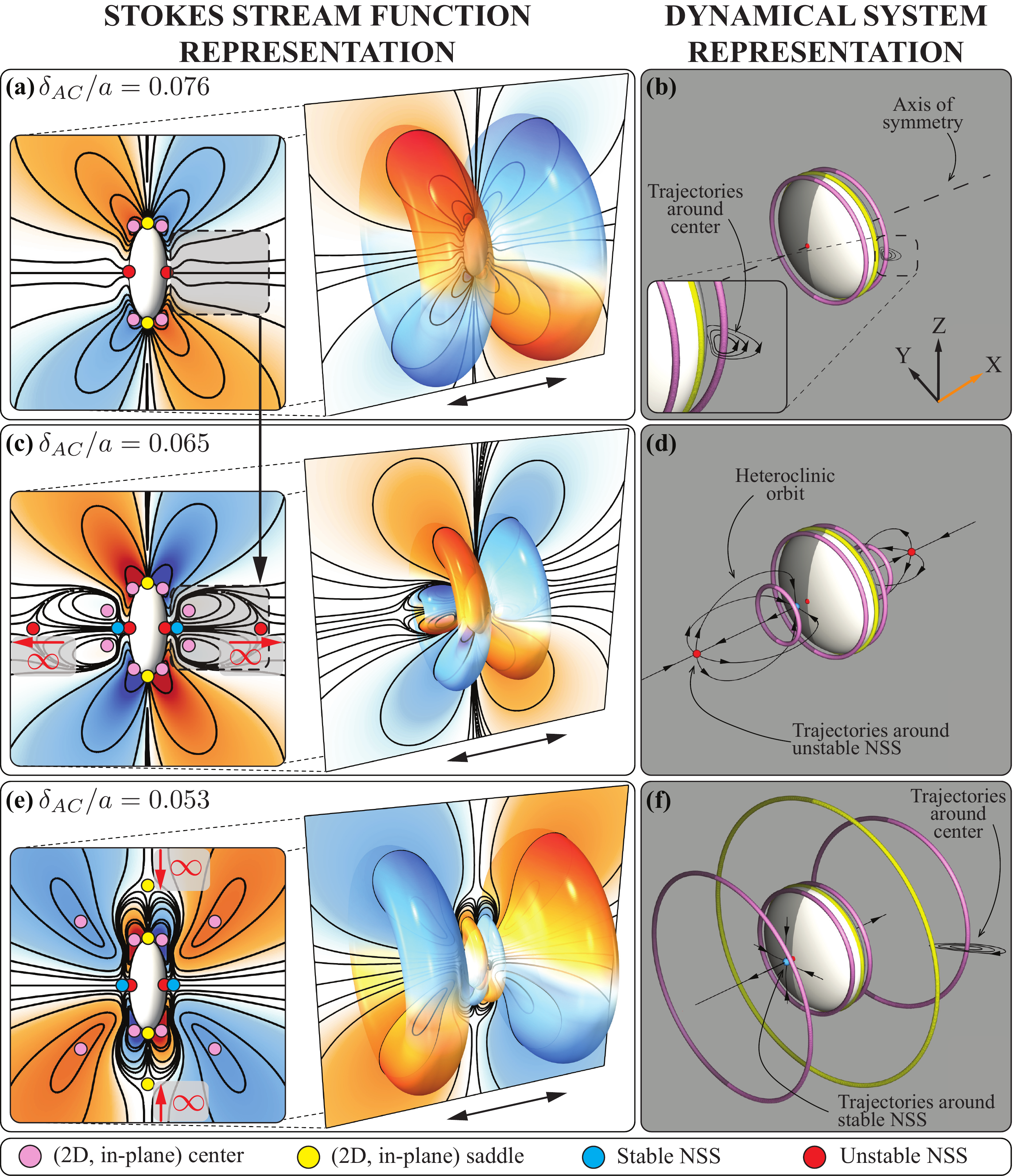}
        \caption{Streaming from an axisymmetric body (spheroid of radii $a_x : a_y : a_z$ = $0.4a : a : a$, where $a = 0.05$): starting from the \textbf{(a,b)} Stokes-like regime ($\delta_{AC} / a = 0.076$), decreasing $\delta_{AC} / a$ transitions the flow topology into \textbf{(c,d)} a new phase ($\delta_{AC} / a = 0.065$) where regions of recirculating fluid are encapsulated between pairs of stable--unstable NSS (shaded boxes). Further decrease in $\delta_{AC} / a$ transitions the flow topology into the \textbf{(e,f)} finite thickness DC layer regime ($\delta_{AC} / a = 0.053$).
        Flow fields are represented using two different methods: Stokes stream function (left column) and dynamical system representation (right column). Simulation details: normalized uniform grid spacing $h / a = 0.02$.}
        \label{fig:axisymmetric}
    \end{center}
\end{figure}

\subsection{Axisymmetric body: Spheroid}
Following a fully symmetric body, we morph the sphere into a spheroid with axis of symmetry aligned with the oscillation direction, thus introducing multiple curvatures while retaining flow axisymmetry.
We consider a spheroid of radii $a_x = 0.4a$ and $a_y = a_z = a$ oscillating along the X-axis (axes defined at the bottom-right of \cref{fig:axisymmetric}(b)).
We note that the two-dimensional equivalent of this system is an oscillating ellipse.
The latter has been previously shown \citep{Bhosale:2020} to give rise to a new flow regime, not attainable by circular cylinders, between the single- (Stokes-like) and double- (finite thickness) layer regimes.
This new flow topology, characterized by closed recirculating pockets of fluid on both sides of the ellipse, can be accessed in 2D either by varying $a_x / a < 1$ at constant $\delta_{AC} / a$, or by fixing $a_x / a < 1$ and changing $\delta_{AC} / a$.
Here we seek confirmation of this behavior in a 3D context, by systematically spanning $\delta_{AC} / a$.

We start by considering high $\delta_{AC} / a$ (low $R_s$), where we encounter the Stokes-like regime. The associated Stokes stream function is illustrated in \cref{fig:axisymmetric}(a), and while the flow is distorted relative to the case of the sphere (\cref{fig:symmetric}(a)), on account of the modified shape geometry, topologically they are equivalent, as confirmed by the dynamical representation of \cref{fig:axisymmetric}(b).
Indeed, we can recognize similar structures, whereby 2D degenerate centers and saddles make up the rings around which the single-layer flow organizes.

Upon decreasing $\delta_{AC} / a$ to a critical value, we observe that the lateral horizontal streamlines (highlighted in \cref{fig:axisymmetric}(a)) are vertically pulled apart and split, locally producing two degenerate centers, a stable (blue) and an unstable (red) NSS that together give rise to neatly enclosed pockets of fluid on both sides of the body (\cref{fig:axisymmetric}(c)).
In the 3D dynamical representation (\cref{fig:axisymmetric}(d)), these structures manifest as outer rings contained within the heteroclinic orbits that connect the stable NSS to the unstable ones, collectively forming the surfaces that separate the fluid within the pockets from the external flow.
Thus, in keeping with 2D observations, a new intermediate flow regime---unattainable in spheres---is identified in 3D.
Such regime is found to form through mechanisms consistent with 2D explanations.
Indeed, the simultaneous appearance of two new centers and two new saddles in the absence of pre-existing critical points is the hallmark of a hyperbolic reflecting umbilic bifurcation \citep{Bosschaert:2013}, as identified in 2D in \cite{Bhosale:2020}.

Thus, by varying flow inertia, the system can be forced to bifurcate, injecting additional topological elements (critical points) that cause the flow to reorganize around newly formed lateral and sealed recirculating regions. Such pockets can then be of practical utility as they provide a mechanism at intermediate flow inertia regimes to, for example, trap, concentrate, manipulate and eventually release microparticles \citep{Parthasarathy:2019}.

Finally, at low $\delta_{AC} / a$ (high $R_s$), we encounter the finite thickness layer regime.
As we decrease $\delta_{AC} / a$, the unstable NSS (red) move away from the body along the axis of oscillation (\cref{fig:axisymmetric}(c)), thus unfolding the pockets.
Eventually, at a critical $\delta_{AC} / a$, the unstable NSS diverge to infinity, opening up the flow laterally.
Concurrently, new 2D degenerate saddles (yellow) approach the body radially from infinity (Supplementary Information) within the YZ-plane (\cref{fig:axisymmetric}(e,f)), ultimately sealing the DC layer by means of heteroclinic orbit connections with the stable NSS (blue).
These degenerate saddles make up the outer yellow-ring of \cref{fig:axisymmetric}(f), leading to a flow topology equivalent to the classic double-layer structure of \cref{fig:symmetric}(d).

We note that the same set of bifurcations and flow regimes, here captured by fixing the spheroid geometry $a_x / a$ and modifying $\delta_{AC} / a$, can be alternatively obtained upon variations on $a_x / a < 1$ at constant $\delta_{AC} / a$ (Supplementary Information), consistent with 2D predictions \citep{Bhosale:2020}.

\section{Non-axisymmetric flows}\label{sec:3d}
\begin{figure}
\begin{center}
\includegraphics[width=0.95\linewidth]{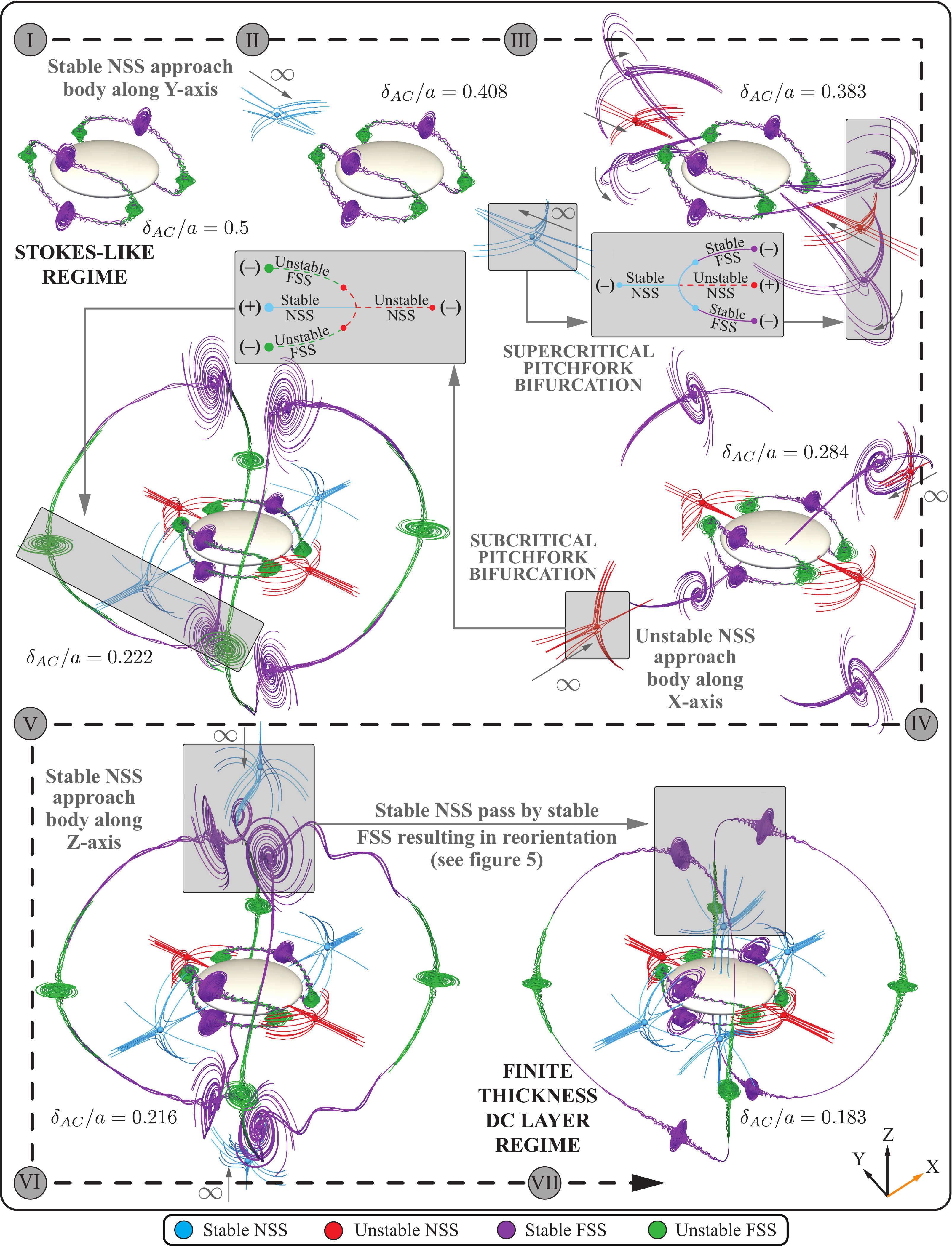}
\caption{Evolution of streaming flow topology for an oscillating spheroid (with radii ratio $a_x : a_y : a_z$ = $a : a : 0.25a$, where $a = 0.05$). We observe seven distinct phases, classified based on critical points and local flow trajectories, from Stokes-like regime (Phase I) to finite thickness layer regime (Phase VII).
Bifurcation diagrams in the insets summarize the transition from one phase to another, tracking stable/unstable (solid/dashed lines) branches as well as type of involved critical points, as a function of the bifurcation parameter $\delta_{AC} / a$.
Axes are indicated at the bottom-right, where the orange X-axis represents the axis of oscillation. Simulation details: normalized uniform grid spacing $h / a = 0.03$ (used throughout this section).}
\label{fig:disc-overview}
\end{center}
\end{figure}

We proceed to investigate shape curvature variation effects in a fully three-dimensional (i.e. non-axisymmetric) setting (\cref{fig:disc-overview}).
We achieve this by considering a spheroid characterized by an inverse aspect ratio ($a_x = a_y = a$, $a_z = 0.25 a$) relative to the case considered above.
This is equivalent to flipping the spheroid of \cref{fig:axisymmetric} horizontally, thus rendering the axis of oscillation (X-axis) perpendicular to the object's axis of symmetry (Z-axis).
Since in this setup the flow is no longer axisymmetric, the Stokes stream function is not available and our analysis can only rely on a dynamical representation, underscoring its utility.
For physical intuition, we henceforth highlight orbits and local flow features by means of passive tracers whose trajectories are colored based on the type of the critical point in the neighbourhood of which they are seeded.
For example, if particles are seeded in the vicinity of a stable NSS (blue), then the corresponding trajectories will be blue.

When spanning flow conditions from high to low $\delta_{AC} / a$ (low to high $R_s$), we observe a rich dynamic behavior.
\Cref{fig:disc-overview} provides an overview of the system's evolution, transitioning from single- (Stokes-like) to double- (finite thickness) layer regime over seven topologically distinct phases described in the following.

\subsection{Phase I $\rightarrow$ II $\rightarrow$ III}\label{subsec:I2III}
We begin by considering high $\delta_{AC} / a$, where the single-layer regime (Phase I) is usually encountered.
In non-axisymmetric, fully 3D settings, degenerate centers and saddles no longer exist, so that the single layer regime manifests in a topologically distinct form.
Indeed, we observe (Phase I) that the rings made of degenerate critical points in \cref{fig:symmetric}(b) and \ref{fig:axisymmetric}(b) are replaced by new ring-like structures, made of four critical points---two stable (purple) and two unstable (green) FSS---connected by heteroclinic orbits (\cref{fig:disc-overview}).
These new rings are effectively the 3D counterpart of the degenerate rings previously discussed, and similarly constitute the skeleton around which recirculating flow regions organize.

As we decrease $\delta_{AC} / a$, we observe that a pair of stable NSS (blue) first approaches the body from infinity (Supplementary Information) along the Y-axis (Phase II), and are subsequently replaced (Phase III) by an unstable NSS (red) and a pair of stable FSS (purple), on both sides of the body.
This Phase II $\rightarrow$ III transition is the result of a two-step process, for which first the stable NSS undergoes a supercritical pitchfork bifurcation \citep{Strogatz:2018} and gives rise to an unstable NSS (red) and a pair of stable NSS (blue), followed by a change in nature of the new stable NSS (blue) into a stable FSS (purple).
This mechanism is confirmed by examining the eigenvalues of the involved critical points,
relative to the direction along which the bifurcation occurs (i.e. Z-direction; X- and Y-directions are sign-invariant throughout the process).
This reveals first a sign change in the real components from $(-) \rightarrow (-,+,-)$ which, in the absence of imaginary parts, denotes the transition from stable NSS to stable NSS, unstable NSS and stable NSS, respectively.
Following the bifurcation, we observe that the eigenvalues of the stable NSS begin to develop imaginary components, marking the initiation of a rotational local flow, thus a change in type from NSS to FSS.
In \cref{fig:disc-overview}, this is illustrated as a bifurcation diagram where we indicate the nature of the critical point (NSS/FSS) and the stability of the branches along which they lie, as we vary the bifurcation parameter $\delta_{AC} / a$.
A full illustration of this two-step process can be found in the Supplementary Information.

\subsection{Phase III $\rightarrow$ IV $\rightarrow$ V}\label{subsec:III2V}
In Phase III, a further decrease in $\delta_{AC} / a$ draws the two unstable NSS (red) closer towards the body along the Y-axis, and pushes the adjacent pairs of stable FSS (purple) farther apart from each other in the YZ-plane, as shown in \cref{fig:disc-overview}.
At a critical $\delta_{AC} / a$ value, a pair of unstable NSS (red) appears (Phase IV) along the X-axis from infinity (Supplementary Information).
This sets the stage for the formation of the outer ring structures eventually expected in the finite thickness layer regime.

When considering the Phase IV $\rightarrow$ V transition, we observe that the two new unstable NSS (red) each splits into a stable NSS (blue) and a pair of unstable FSS (green), through a two-step process similar to Phase II $\rightarrow$ III, except that the transition here is mediated by a subcritical pitchfork bifurcation \citep{Strogatz:2018}.
The appearance of these unstable FSS causes a drastic remodeling of the flow.
Indeed, the simultaneous presence of unstable FSS (green) in the XY-plane and of stable FSS (purple) in the YZ-plane forces the flow to form heteroclinic connections which altogether define a pair of outer ring structures.
Nonetheless, this flow topology does not correspond to the classic double-layer regime yet: indeed, the outer rings are orthogonal to the inner ones!
This makes up a complex flow structure for which inner and outer regions are characterized by perpendicular crossflows.

\begin{figure}
\begin{center}
\includegraphics[width=\linewidth]{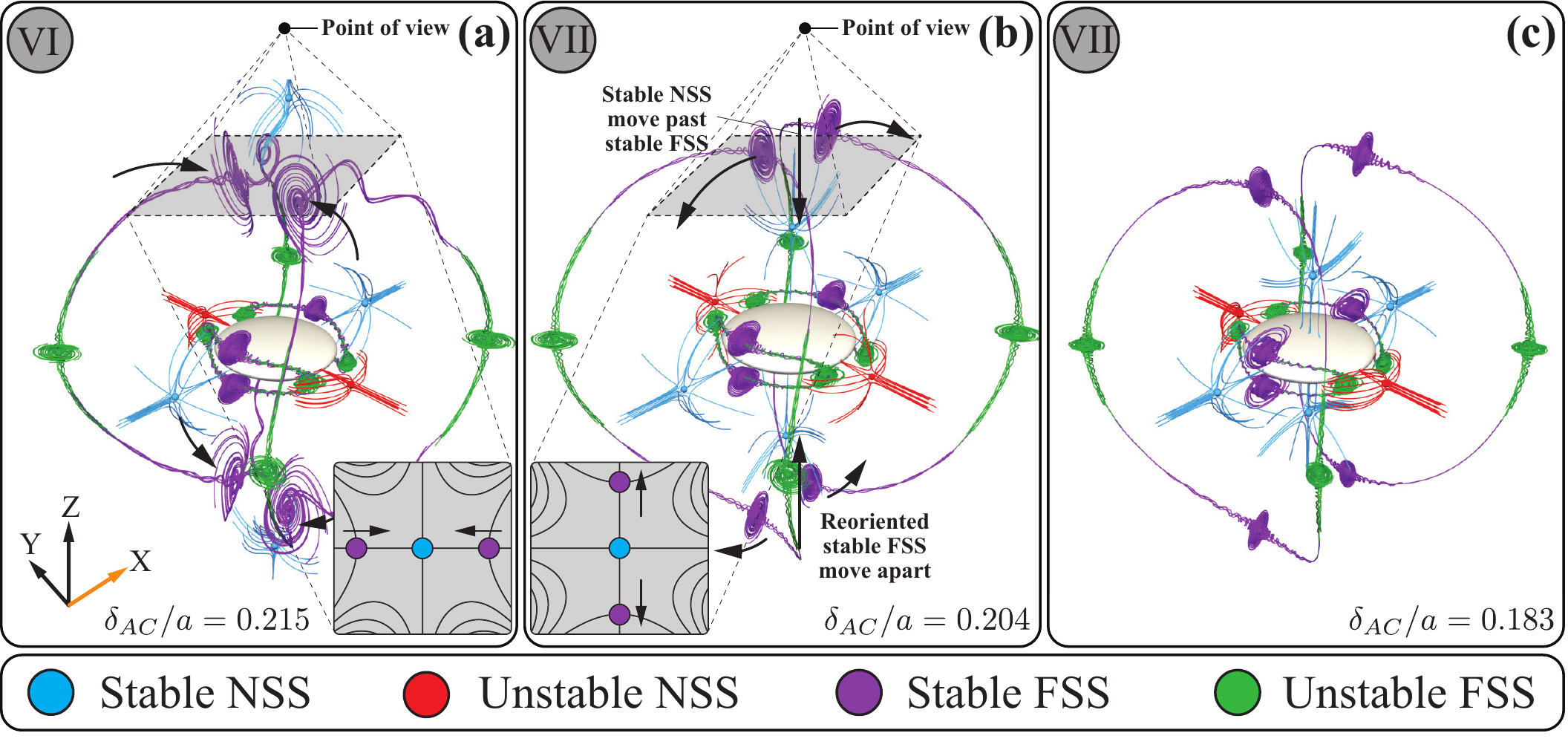}
\caption{Phase VI $\rightarrow$ VII:
Streaming flow structures \textbf{(a)} before and \textbf{(b)} after transition. The corresponding steady state flow trajectories on the highlighted planes are shown in the insets, illustrating critical points coming together one on axis, and splitting away on another---a feature characteristic of higher order elliptic reflecting umbilic bifurcations \citep{Bosschaert:2013}.
\textbf{(c)} Flow converges to the finite thickness layer regime upon further decrease in $\delta_{AC} / a$.}
\label{fig:disc-vi2vii}
\end{center}
\end{figure}

\subsection{Phase V $\rightarrow$ VI $\rightarrow$ VII}\label{subsec:V2VII}
Finally, as we further decrease $\delta_{AC} / a$, we recover the expected double-layer regime.
From Phase V, we first observe the appearance of a new pair of stable NSS (blue) approaching from infinity along the Z-axis (Phase VI).
These are drawn towards the body and thus towards the pairs of stable FSS (purple) in the outer rings.
In doing so, the stable NSS (blue) deform the outer rings inwards, causing top and bottom FSS (purple) to come closer together.
Eventually, the stable NSS (blue) pass through the FSS (purple) pairs, at which point the outer rings ``kiss'' and reorient orthogonally, reorganizing the flow into the double layer regime of Phase VII.

This qualitative dynamic portrait can be rigorously analyzed in terms of bifurcations by projecting the involved critical points on the grey planes, parallel to the XY-plane, illustrated in \cref{fig:disc-vi2vii}.
We can observe (insets) how the two stable FSS approach each other along the Y-axis, collide with the stable NSS, and then move away from one another along the X-axis.
In this characteristic orthogonal rearrangement of critical points, we can recognize a higher-order elliptic reflecting umbilic bifurcation \citep{Bosschaert:2013} at work.
As a consequence, the heteroclinic orbits between the stable and unstable FSS in the outer rings break up and orthogonally reconnect, forming new rings that are now consistently oriented with the inner ones.
This final topology can be even more clearly appreciated as we further decrease $\delta_{AC} / a$ in \cref{fig:disc-vi2vii}(c).
We see that the outer rings make up the core of a recirculating flow region that extends from infinity down to the stable/unstable NSS (blue/red).
These in turn, together with their connecting orbits, define the surface that separates outer from inner flows, with the latter recirculating around the inner rings, tightly fit to the body.
The overall flow architecture of Phase VII is then found to be consistent with the finite thickness layer regime of \cref{fig:symmetric}(c,d) and \ref{fig:axisymmetric}(e,f).

Finally, we note that the flow topological rearrangements observed in the investigations above can also be achieved via geometrical variations alone (i.e. by changing $a_z / a < 1$ while keeping $\delta_{AC} / a$ constant), as demonstrated in the Supplementary Information and consistent with the axisymmetric case of \cref{fig:axisymmetric}.

\section{Topologically distinct body}\label{sec:topology}
\begin{figure}
\begin{center}
\includegraphics[width=\linewidth]{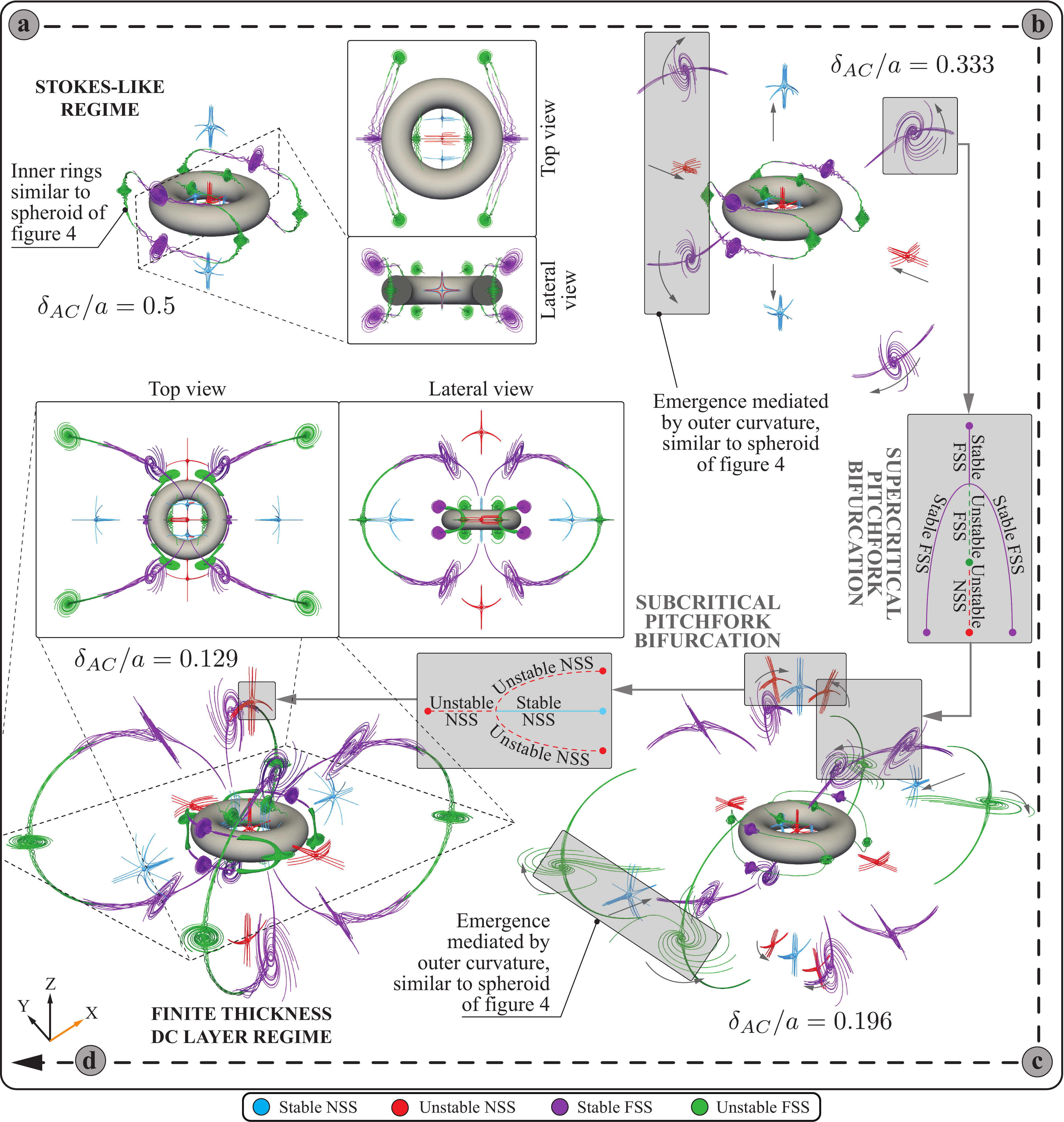}
\caption{Streaming flow topology at different $\delta_{AC} / a$ for a torus, with axis of symmetry aligned along the Z-axis, oscillating along the X-axis.
The torus has tube radius $a_{tube} = 0.25 a$ and core radius $a_{core} = 0.75 a$, so that the overall body length scale is $a = a_{tube} + a_{core}$. We highlight here that the torus has major ($a_{max} = a$) and minor ($a_{min} = a_{tube} = 0.25 a$) length scales comparable to that of the three-dimensional spheroid in \Cref{sec:3d}, essentially presenting equivalent width and thickness.
Simulation details: normalized uniform grid spacing $h / a = 0.03$.}
\label{fig:torus}
\end{center}
\end{figure}

After investigating streaming flows in terms of body geometry and flow inertia variations, we finally begin to probe the effects of shape topological changes.
The inextricable connection between topology and geometry provides a vast manipulation space that can hardly be systematically explored.
Hence, here we narrow down the scope of our investigation and consider a single topological defect---a hole---in a spheroid similar to \cref{fig:disc-overview}.
We thus transition from a genus-0 spheroid (body with no holes) to a genus-1 torus (body with one hole) of comparable length scales.
We then consider four representative $\delta_{AC} / a$ values between the Stokes-like and the finite thickness DC layer regime, as shown in \cref{fig:torus}.
This gives us the opportunity to begin to understand how flow structures pertinent to the spheroid remodel due to the interaction with the flow within the topological defect.

When considering high $\delta_{AC}/ a$, we encounter the streaming flow topology representative of the Stokes-like regime as depicted in \cref{fig:torus}(a).
In this regime, we observe the presence of two rings, fit to the body and made of two stable (purple) and unstable (green) FSS connected by heteroclinic orbits, similar to those encountered in \cref{fig:disc-overview}.
These rings are unaffected by the topological defect.
Indeed, due to the proximity to the body, the flow effectively detects only the object's outer curvature, which is similar to the spheroid.
Within the topological defect, we observe a collection of critical points (one unstable NSS (red), two stable NSS (blue) and four unstable FSS (green)---inset).
Additionally, a pair of stable NSS are found along the Z-axis, off the XY-plane.
This particular pair, as we will see, plays a significant role in remodeling the flow relative to the genus-0 spheroid.
Indeed, they extend the influence zone of the topological defect, providing opportunities for structures exterior to the torus to eventually interact.

As we decrease $\delta_{AC} / a$, we see an interesting mechanism play out. At first the flow exterior to the torus evolves in accordance to the spheroid case (Phase I $\rightarrow$ III), whereby two stable NSS approach from infinity along the Y-axis, and then undergo a pitchfork bifurcation forming the unstable NSS (red) and pairs of stable FSS (purple), on both sides of the torus (\cref{fig:torus}(b)).
This process is thus still governed by the object's outer curvature.
Nonetheless, after the pitchfork bifurcation takes place, the pair of purple FSS progressively fan out in the YZ-plane and approach the defect's zone of influence, represented by the two stable NSS along the Z-axis.
The result of this interaction manifests in the first departure from the flow evolution depicted in \cref{fig:disc-overview}.
Indeed, the stable FSS (purple) now further bifurcate into a pair of stable FSS and an unstable NSS (\cref{fig:torus}(c)).
We note that here we encounter again a two-step mechanism: first we have a supercritical pitchfork bifurcation (stable FSS $\rightarrow$ 2 stable FSS + 1 unstable FSS) followed by a change in nature from FSS to NSS on the unstable branch.
Concurrently, on the XY-plane we observe the appearance of the two unstable FSS (green) and stable NSS (blue) already seen in Phase III $\rightarrow$ V of the spheroid (\cref{fig:disc-overview}, \cref{subsec:III2V}).
This is consistent with the intuition that external flow structures, especially in the XY-plane, are insensitive to the topological defect and primarily respond to the object's outer curvature.
Overall, this process sets the stage for a dramatic reconfiguration of the finite thickness regime, relative to the spheroid.
Indeed, in the outer flow region, there are now eight stable and only four unstable FSS.
Thus, these critical points no longer have the opportunity to form the pair of outer rings (each made of two stable and two unstable FSS) of \cref{fig:disc-overview}.
Instead, they are forced to connect in a new structure capable of accommodating the four extra stable FSS.
The solution is offered by a clover-like ring structure running through the midpoint of the topological defect (\cref{fig:torus}(d)).
This meeting point has also the effect of ``locking'' the rings in, preventing any further re-orientation, unlike those observed for the spheroid (Phase V and VI in \cref{fig:disc-overview}).
The resulting flow now fundamentally differs from previously observed finite thickness regimes.
In fact, although we can still identify a DC layer organized around the smaller rings fit to the body, this recirculating flow region is now confined by an outer flow that both extends to infinity and permeates the center of the domain by merging through the hole of the torus.
This unique configuration may offer novel microfluidic opportunities, whereby the easily accessible outer flow provides now a natural mechanism to transport particles from the top and bottom of the torus to the topological defect, thus focusing them for self-assembly or mixing applications.

\section{Conclusion}\label{sec:conclusion}
Towards the goal of extending our understanding of streaming flow dynamics in three dimensional settings, we start by revisiting the classical case of the oscillating sphere and present observed flow structures and transitions through the lens of dynamical systems theory \citep{Bhosale:2020}.
We further demonstrate the utility and extensibility of this approach to understand streaming flows in more general, but still axisymmetric 3D cases.
We then systematically investigate streaming in a fully three-dimensional setting by oscillating a spheroid perpendicular to its axis of symmetry, revealing a rich dynamic behavior which we understand using bifurcation theory.
Finally, we present a first foray into streaming induced by a topologically distinct body.
Thus, a torus of length scales comparable with the previously investigated spheroid is analyzed, revealing intriguing flow organizations of potential utility for microparticle concentration, self-assembly and mixing.
Altogether, these results provide physical intuition, principles and analysis tools to manipulate 3D streaming flows based on body geometry, topology and flow inertia, with potential applications in microfluidics and micro-robotics.


\section{Acknowledgements}\label{sec:ack}
The authors acknowledge support by the National Science Foundation under NSF CAREER Grant No. CBET-1846752 (MG) and by the Blue Waters project (OCI-0725070, ACI-1238993), a joint effort of the University of Illinois at Urbana-Champaign and its National Center for Supercomputing Applications.
This work also used the Extreme Science and Engineering Discovery Environment (XSEDE) \citep{Towns:2014} Stampede2, supported by National Science Foundation grant no. ACI-1548562, at the Texas Advanced Computing Center (TACC) through allocation TG-MCB190004.
We thank S. Hilgenfeldt for helpful discussions over the course of this work and Wim M. van Rees for technical support on three-dimensional simulations.


\bibliographystyle{unsrtnat}

\end{document}